\documentclass[fleqn,preprint,showpacs,preprintnumbers,amsmath,amssymb]{revtex4}
\usepackage{graphicx}
\usepackage{dcolumn}
\usepackage{bm}

\begin{document}

\title{Coulomb Stability-Shatter: Consequences for Compact-Stars}
\author{M. Akbari-Moghanjoughi}
\affiliation{Azarbaijan University of
Tarbiat Moallem, Faculty of Sciences,
Department of Physics, 51745-406, Tabriz, Iran}

\date{\today}
\begin{abstract}
In this research the transverse hydrostatic stability of a gravitating Fermi-Dirac plasma under a quantizing field is explored. It is revealed that such plasma is magnetically unstable due to the Landau electron spin-orbit quantization in all magnetic levels which might explain the gravitational compactification in the case of magnetic compact-stars. It is also revealed that the Coulomb interaction plays a fundamental role in the magnetic instability when a critical field is reached for a given plasma composition. It is remarked that, the Coulomb interaction, on specific condition depending on the field-strength and plasma composition, can suddenly destroy the transverse stability of the relativistically degenerate Fermi-Dirac plasma. The consequences of such instability on mass-limit of magnetic compact-stars and the neutron star crust with possible relation to gamma-ray bursts is numerically examined.
\end{abstract}

\maketitle

\section{Introduction}

Astrophysical compact plasma is a unique environment in which newtonian gravity, special relativity, quantum mechanics and electromagnetism all play equal parts. Perhaps the most distinguished exhibition of large-scale classical gravity and quantum mechanics collaboration is the effect of relativistic degeneracy of Fermi-Dirac electrons leading to the famous Chandrasekhar mass-limit on compact-stars called white-dwarfs \cite{chandra}. It has been shown that the equation of state \cite{kothary} as well as nonlinear wave dynamics \cite{akbari1} of a degenerate plasma is affected fundamentally by quantum degeneracy when the de Broglie thermal wavelength $\lambda_B = h/(2\pi m_e k_B T)^{1/2}$ becomes comparable or higher than inter-fermion distances \cite{landau}. There are other effects such as Thomas-Fermi crystallization, Coulomb interaction, electron exchange and ion correlation effects which might contribute to stability of quantum plasmas. However, it has been shown \cite{salpeter} that these effects give-rise to correction-terms of order the fine-structure constant or even smaller to the hydrostatic stability of gravitating relativistically degenerate plasmas slightly altering the mass-radius relation only for high atomic-number composed compact stars \cite{suh}.

One of the greatest challenges in astrophysical sciences is to understand puzzling presence and possible role of magnetism in stellar birth \cite{fermi, attard} collapse, \cite{weeler}, evolution \cite{ginz, wolt} and physical processes such as gamma-ray bursts \cite{max, duk}. There have been reports of white-dwarfs and neutron stars with associated ultra-high magnetic fields \cite{crut, gold}. Extensive past studies \cite{can1, can2, can3} reveal that the magnetism can lead to quantization of thermodynamic quantities and electronic density of state in Fermi-Dirac gas. The magnetic instability has been shown to dominate due to anisotropic stress in ground-state Landau-level \cite{can4}. The oscillatory feature of spin-orbit magnetization in super-dense plasma may also lead to a metastable self-sustained ferromagnetism (called Landau orbital ferromagnetism or LOFER) with field strength as high as $10^8G$ and $10^{13}G$ for white-dwarf and neutron-star density ranges \cite{lee, con}. However, the LOFER state has been shown to disappear at temperatures estimated for such compact-stars \cite{burk}. Recently, it has been shown that the magnetic transverse collapse due to pressure anisotropy in magnetized neutron-star \cite{chai} and LOFER plasma \cite{akbari2} is possible.

One of important astrophysical motivation towards the study of matter under extreme conditions such as in strong magnetic-field and relativistic degeneracy arises due to importance of understanding the neutron star crust which play fundamental role in many astrophysical processes and observed phenomena such as gamma-ray bursts \cite{dong}. The study of matter under strong magnetic field has also implication for the strong laser-matter interaction experiments \cite{mend}. Many models has been proposed for the underlying mechanism for gamma-ray bursts \cite{tsvi}. However, the inner engine for such gigantic energy release is far from clear. More recently, a model of resonant excitation in the neutron star (NS) crust has been proposed \cite{tsang}. It has been shown that the crust-core interaction may lead to cracks in the crust leading to energy release of order $10^{46}ergs$. However, the crust tension in this and many other previous models rely on the external mergers such as that of NS-NS or NS-black hole. In the present paper we show that the instability and shatter of a neutron-star crust is possible due to Landau quantization accompanied by Coulomb interactions without a need for external cause.

\section{Stability Criterion}\label{stability}

Considering a uniformly magnetized homogenous gravitating Fermi-Dirac quasi-neutral plasma, the hydrostatic stability is defined through the force equilibrium, $\bf{F}_{in}=\bf{F}_{out}$ on an spherical shell element, $dr$, may be written as
\begin{equation}\label{eq}
\frac{1}{\rho }\frac{{dP}}{{dr}} =  - \frac{{GM(r)}}{{{r^2}}},
\end{equation}
which at the quasi-neutral state, i.e. $n_e\simeq n_i$, the local center of mass plasma density will be $\rho=m_i n_i + m_e n_e\simeq m_i n_e$. The local pressure of relativistically-degenerate electrons compose almost the total pressure which can be anisotropic, $P_{\parallel}=P_e=P_{\perp}+\Gamma B$, in the presence of a uniform magnetic field, in which $\Gamma$ is the magnetization. The electron pressure and density are Landau-quantized and at the zero-temperature approximation take the form \cite{can1}
\begin{equation}\label{nd}
\begin{array}{l}
{\rho}({\epsilon _{Fe}},\gamma ) = {\rho_c}\gamma \sum\limits_{l = 0}^{{l_m}} {(2 - {\delta _{l,0}})} \sqrt {\epsilon _{Fe}^2 - 2l\gamma - 1},\\
{P_{\parallel}}({\epsilon _{Fe}},\gamma ) = \frac{1}{2}{\rho_c}\gamma {m_e}{c^2}\sum\limits_{l = 0}^{{l_m}} {(2 - {\delta _{l,0}})} \left\{ {{\epsilon _{Fe}}\sqrt {\epsilon _{Fe}^2 - 2l\gamma  - 1} } \right. \\ \left. { - (1 + 2l\gamma )\ln \left[ {\frac{{{\epsilon _{Fe}} + \sqrt {\epsilon _{Fe}^2 - 2l\gamma  - 1} }}{{\sqrt {2l\gamma  + 1} }}} \right]} \right\},\\ \Gamma({\epsilon _{Fe}},\gamma ) = B_c^{-1}\frac{{\partial {P_{\parallel}}({\epsilon _{Fe}},\gamma )}}{{\partial \gamma}}.
\end{array}
\end{equation}
where, $\rho_c=m_im_e^3 c^3/2\pi^2 \hbar^3$, $\gamma=B/B_c$ with $B_c=m_e^2c^3/e\hbar\simeq 4.414\times 10^{13}G$. The parameter $\epsilon _{Fe}=\sqrt{1+x^2}$ is the normalized relativistic Fermi-energy which is related to the relativity parameter, $x=p_{Fe}/m_e c$, the normalized Fermi electron-momentum. The maximum occupied Landau-level, $l_m$, is defined as $l_m=(\epsilon _{Fe}^2-1)/2\gamma\geq l$ and the parameter $\delta_{l,0}$ takes values of $\{1,0\}$, respectively, for the quantum-limit ($l=0$) and for other $l$-values. The corresponding Chandrasekhar equation-of-state may be obtained from Eq. (\ref{nd}) as $l \rightarrow \infty$. The transverse hydrostatic stability can be written as
\begin{equation}\label{eq1}
\frac{1}{{4\pi {r^2}G}}\frac{d}{{dr}}\left( {\frac{{{r^2}}}{{\rho (x,\gamma )}}\frac{{\partial \left[ {{P_\parallel }(x,\gamma ) - B\Gamma (x,\gamma )} \right]}}{{\partial r}}} \right) + \rho (x,\gamma ) = 0,
\end{equation}
On the other hand, the equation of state of a magnetized Fermi gas can be evaluated analytically in terms of Hurwitz functions \cite{claud} in the following form
\begin{equation}\label{h}
\begin{array}{l}
\rho(x,\gamma ) =\rho_c{(2\gamma )^{3/2}}{H_{ - 1/2}}\left( {\frac{{{x^2}}}{{2\gamma }}} \right), \\
{P_{\parallel}}(x,\gamma ) = \frac{{1}}{{2}}c_{s}^{2}{\rho _c}{(2\gamma )^{5/2}}\int_{0}^{\frac{{{x^2}}}{{2\gamma }}} {\frac{{{H_{ - 1/2}}(q)}}{{\sqrt {1 + 2\gamma q} }}} dq,\\
{H_p}(q) = h(p,\{ q\} ) - h(p, q + 1 ) - \frac{1}{2}{q^{ - p}}, \\
h(p,q) = \sum\limits_{n = 0}^\infty  {{{(n + q)}^{ - p}}} . \\
\end{array}
\end{equation}
where $h(p,\{q\})$ is the Hurwitz zeta-function of order $p$ with the fractional part of $q$ as argument and $c_s=c\sqrt{m_e/m_i}$ is the wave-speed. Using the definitions $\rho({\epsilon _{Fe}},\gamma ) = c_{s}^{-2}\partial {P_\parallel({\epsilon _{Fe}},\gamma )}/\partial {\epsilon _{Fe}}$ and $\Gamma({\epsilon _{Fe}},\gamma ) = B_c^{-1}{{\partial {P_{\parallel}}({\epsilon _{Fe}},\gamma )}}/{{\partial \gamma}}$ for a homogenous magnetized plasma case, we have
\begin{equation}\label{eq2}
\frac{1}{{\rho (x,\gamma )}}\frac{{\partial \left[ {{P_\parallel }(x,\gamma ) - \Gamma (x,\gamma )B} \right]}}{{\partial r}} = \frac{{c_{s}^2x(r)}}{{\sqrt {1 + x{{(r)}^2}} }}\left[ {1 - \frac{\gamma }{{\rho (x,\gamma )}}\frac{{\partial \rho (x,\gamma )}}{{\partial \gamma }}} \right]\frac{{dx(r)}}{{dr}},
\end{equation}
which, after some algebra, leads to the general stability relation below for the uniformly magnetized gravitating Fermi-Dirac plasma
\begin{equation}\label{eq3}
\frac{{c_{s}^2}}{{4\pi {r^2}G}}\frac{d}{{dr}}\left\{ {\frac{{{r^2}x(r)}}{{\sqrt {1 + x{{(r)}^2}} }}\left[ {1 - \frac{\gamma }{{\rho (x,\gamma )}}\frac{{\partial \rho (x,\gamma )}}{{\partial \gamma }}} \right]\frac{{dx(r)}}{{dr}}} \right\} + \rho (x,\gamma ) = 0.
\end{equation}
The parameter $\Omega(x,\gamma)=\gamma\partial\ln \rho(x,\gamma)/\partial \gamma$ defines the stable solution to Eq. (\ref{eq3}).  Figure 1 shows the variation of mass-density (along with the monotonic and oscillatory components) and stability parameter $\Omega(x,\gamma)$ in terms of the relativity parameter, $x$, indicating that the stability solution can become unstable for some discrete $x$-ranges for a given magnetic field parameter, $\gamma$. The linear part of Fig. 1(a) corresponds to the quantum-limit ($l=0$) for which $\rho(x,\gamma)=\gamma x$. On the other hand, at the weak quantization limit ($l=\infty$) the magnetization vanishes and we have $\rho=x^3$, where, the plasma becomes magnetically stable and isotropic in all directions. The corresponding Landau-level is shown in Fig. 1(b) to be critically-unstable ($\Omega(x,\gamma)=1$). Landau-levels along with the corresponding quantized stability regions are depicted in Fig. 2. The white regions in Fig. 2(b) indicate the stability voids for which $\Omega(x,\gamma)\leq 1$. For the quantum-limit, $\Omega(x,\gamma)= 1$, (the largest white-region to the left in Fig. 2(b)) the plasma can free-fall under its own gravity, while, for other Landau-levels the total pressure becomes negative and the plasma collapse may occur even in the absence of gravity.

\section{Coulomb Effect}\label{coulomb}

It has been shown that the effects such Thomas-Fermi screening, Coulomb attraction, exchange and correlation effects can affect the stability of the relativistically degenerate plasma \cite{salpeter}. The former three effects are of the order of fine-structure constant, while the correlation effect is much smaller. Here, we consider the Coulomb effect which can be dominant particularly in the presence of a magnetic field, since, the degeneracy of the plasma can be dominated by the new bound-states introduced by Landau quantization \cite{dong}. The attractive Coulomb interaction introduces another negative term to the pressure of plasma, as below \cite{suh}
\begin{equation}\label{c}
{P_C}(x,\gamma) =  - \frac{{18{\pi ^2}{m_e^4}{c^5}}}{{{h^3}}}\left( {\frac{{{\alpha ^5}{z^{2/3}}}}{{10{\eta^4}}}} \right),\hspace{3mm}\eta = \alpha {\left[ {\frac{{3\pi}}{{8\gamma \rho(x,\gamma)}}} \right]^{1/3}}. \\
\end{equation}
where, the parameters, $\alpha=e^2/\hbar c$ and $z$ are the fine-structure constant and the atomic number, respectively. Hence, in the presence of Coulomb effect the stability parameter can be written as
\begin{equation}\label{sc}
\Omega (x,\gamma,z ) = \frac{\gamma }{{\rho (x,\gamma )}}\frac{{\partial \rho (x,\gamma )}}{{\partial \gamma }} + \frac{{\beta \sqrt {1 + {x^2}} {z^{2/3}}{\gamma ^{4/3}}}}{{x\rho {{(x,\gamma )}^{2/3}}}}\frac{{\partial \rho (x,\gamma )}}{{\partial x}}, \hspace{3mm}\beta  = \frac{{64{\pi ^{2/3}\alpha}}}{{45\times {3^{1/3}}}}.
\end{equation}
where, we have used the identity
\begin{equation}\label{fc}
{F_C} =  - \frac{{1}}{{\rho (x,\gamma )}}\frac{{\partial {P_C}}}{{\partial \rho (x,\gamma )}}\frac{{\partial \rho (x,\gamma )}}{{\partial x}}\frac{{\partial x}}{{\partial {\epsilon _{Fe}}}}\frac{{\partial {\epsilon _{Fe}}}}{{\partial r}}.
\end{equation}
The stability regions for two different plasma compositions, namely carbon and iron, are shown in Fig. 3. It is remarked that for a given critical-value of fractional field-parameter, $\gamma$, the stability is shattered. Surprisingly, the Coulomb stability-shatter for iron composition is observed to coincide with critical-field value, $B\simeq B_c$. Evidently, this effect can lead to shattering in the external iron-rich crust of a critically-magnetized neutron star and allow to a enormous energy release such as gamma-ray bursts. In a recent paper Tsang et al. \cite{} have shown that such energy release is possible due to a resonant shatter in neutron-star crust.

\section{Finite-Temperature Effect}\label{temp}

The finite-temperature effects might be important for some white-dwarfs with typical temperature of order $10^6$ Kelvins. The thermodynamic quantities of finite-temperature may be calculated from the zero-temperature counterparts, through the relation \cite{claud}
\begin{equation}\label{t}
Q(\tau ,\mu ) = \int\limits_{\frac{{1 - \mu }}{\tau }}^\infty  {\frac{{Q(0,\mu  + \xi \tau )}}{{4{{\cosh }^2}(\xi /2)}}d\xi },\hspace{3mm}\tau=\frac{k_B T}{m_e c^2},
\end{equation}
where, $\mu$ is the chemical potential and $Q$ can be an arbitrary thermodynamic quantity. For instance the density $\rho(\mu,\tau,\gamma)$ can be obtained from $\rho(x^{'},\gamma)$ with argument $x^{'}=\sqrt{x^2+2\mu x \tau+\mu^2\tau^2}$, which ignoring the $\mu^2\tau^2$ for small values of $\tau$ leads to
\begin{equation}\label{t}
\rho (\tau ,\mu ,\gamma ) \approx \int\limits_{ - \infty }^\infty  {\frac{{{{(2\gamma )}^{3/2}}}}{{4{{\cosh }^2}(\xi /2)}}{H_{ - 1/2}}\left( {\frac{{{x^2} + 2\mu \tau \xi }}{{2\gamma }}} \right)d\xi }.
\end{equation}
For a comparison Fig. 4 shows the effect of finite-temperature (ignoring the Coulomb interaction effect) on plasma-density. It is observed that the oscillatory part of the thermodynamic quantities introduced by $h(p,\{q\})$ in Eq. (\ref{h}) is smoothed in finite-temperature case, however, the magnitude is not affected significantly. Due to such effect one expects that the stability quantization in Fig. 2(b) to be smeared out at high temperatures. To evaluate the finite-temperature effect on plasma instability one may consider only the monotonic part of density function in Eq. (\ref{h}). Figure 5 shows the stability regions for the above-mentioned compositions taking into account only the monotonic part of density function. It is remarked that, the destabilization effect persists even with monotonic variation of thermodynamic quantities and is not specific to the quantized case. This is an indication of the fact that the Coulomb instability can even be present at finite-temperature case. Also, the values of field-parameters for which the instability starts in Fig. 5 are of the same order of magnitude in Fig. 3.

\section{The Mass-Limit}\label{calculation}

To consider the aforementioned instability on mass-limit we will consider only the monotonic variation and calculate based on the method given in Ref. \cite{gar}. The single differential equation involving the stellar profile transverse to the magnetic-field is of the form
\begin{equation}\label{eqx}
\frac{{3\pi \mathchar'26\mkern-10mu\lambda _c^2{n_h}}}{{16{r^2}}}\frac{d}{{dr}}\left\{ {\frac{{{r^2}x(r)}}{{\sqrt {1 + x{{(r)}^2}} }}\left[ {1 - \Omega (x,\gamma ,z)} \right]\frac{{dx(r)}}{{dr}}} \right\} + {(2\gamma )^{3/2}}{H_{ - 1/2}}\left( {\frac{{x{{(r)}^2}}}{{2\gamma }}} \right) = 0,
\end{equation}
where, the stability parameter $\Omega(x,\gamma,z)$ is defined in Eq. (\ref{sc}) and $n_h=\sqrt{\hbar c/G}/m_i\simeq 1.3\times 10^{19}$ is the dimensionless hierarchy-number defined based on three fundamental quantum, relativity and classical constants, namely, the scaled Plank-constant, $\hbar$, the speed of light in vacuum, $c$ and the gravitational constant, $G$. It is clearly evident that, the hydrostatic stability of the configuration given by Eq. (\ref{eqx}) is lost when the term in the brackets in this equation becomes negative or vanishes. We now move-on to evaluate the stable configuration for given values of relativity and field parameters.

Each stable configuration is found by integrating Eq. (\ref{eqx}) from the center ($r=0$) outwards until $x(r)$ vanishes for some $r=r_s$ which would be the surface (of the star). The initial condition, however, is set to $x(r=0)=x_c$ and $dx/dr(r=0)=0$ such that for every given value of the relativity parameter at the center ($x_c$) there will be a distinct stellar configuration. It is convenient to change the Eq. (\ref{eqx}) to the normalized form by change of variables $y=x/x_c$ and $\xi=r/r_c$, where, $r_c=\sqrt{3\pi}n_h\mathchar'26\mkern-10mu\lambda_c/4x_c$, so that at the center ($\xi=0$) we will have $y=1$. The Eq. (\ref{eqx}), then in dimensionless form, reads as
\begin{equation}\label{eqy}
\frac{1}{{{\xi ^2}}}\frac{d}{{d\xi }}\left\{ {\frac{{{x_c}{\xi ^2}y(\xi )}}{{\sqrt {1 + x_c^2y{{(\xi )}^2}} }}\left[ {1 - \Omega (x,\gamma ,z)} \right]\frac{{dy(\xi )}}{{d\xi }}} \right\} + \frac{{{{(2\gamma )}^{3/2}}}}{{x_c^3}}{H_{ - 1/2}}\left( {\frac{{x_c^2y{{(\xi )}^2}}}{{2\gamma }}} \right) = 0.
\end{equation}
It is remarked (making use of Eq. (\ref{sc})) that, in the field-free limit, the above relation reduces to the famous Lane-Emden equation of index 3 leading to the well-known Chandrasekhar mass-limit of $M_{Ch}\simeq1.43217M_S$ ($M_S$ being the mass of the sun). A standard integration algorithm may be used to find, for instance, the value of $\xi=\xi_s$ at which $y(\xi)$ vanishes (the surface of star) from which the radius and the mass of the stable configuration (star) can be calculated via the following relations
\begin{equation}\label{rm}
{r_s} = {r_c}{\xi _s},\hspace{3mm}M = 4\pi \int_0^{{r_s}} {\rho (r){r^2}dr}  = \frac{3{\sqrt {3\pi } }}{16x_c^{3}}{m_i}n_h^3{(2\gamma )^{3/2}}\int_0^{{\xi _s}} {{H_{ - 1/2}}\left( {\frac{{x_c^2y{{(\xi )}^2}}}{{2\gamma }}} \right){\xi ^2}d\xi }.
\end{equation}
In the next section, for numerical scheme we will use the unique scalings $n_h\mathchar'26\mkern-10mu\lambda_c\simeq 0.787 R_E$ ($R_E$ being the radius of Earth) and $m_i n_h^3\simeq 1.849 M_S$ ($M_S$ being the mass of Sun), which are conventionally employed in literature. Figure 6 shows the solution of $y(\xi)$ of Eq. (\ref{rm}) for different values of fractional field-parameter, $\gamma$, for dense-core configuration ($x_c\simeq \infty$) for uniform iron composition. It is evident that the mass-limit varies according to the change in magnetic field strength. However, the configuration becomes unstable (dashed curves) for values of $\gamma$ for which the stability is shattered (e.g. see Fig. 3). The composition dependent of the normalized mass-limit id depicted for three different compositions in Fig. 7 making a bold conclusion about the well-known Chandrasekhar mass-limit. It should be noted that although the field strength considered here may be vary high compared to those measured for white-dwarfs surface, these values can be realized for neutron stars and even for white-dwarf cores \cite{angel}.

\newpage

\textbf{FIGURE CAPTIONS}

\bigskip

Figure-1

\bigskip

Variation of the normalized plasma mass-density (along with oscillatory and monotonic parts) and stability parameter, $\Omega(x,\gamma)$, with respect to the fractional Fermi-momentum, $x$ (the relativity parameter) for a given magnetic field parameter, $\gamma=0.3$.

\bigskip

Figure-2

\bigskip

Regions of the Landau magnetic-levels and the corresponding regions of stability voids (unstable regions in white) for a uniformly magnetized zero-temperature Fermi-Dirac plasma caused by the spin-orbit quantization.

\bigskip

Figure-3

\bigskip

Coulomb stability-shatter for two different plasma composition, namely, the carbon and iron. The Coulomb stability-shatter starts at a given critical magnetic-field value ($\gamma\simeq$ 2.5 for carbon and $\gamma\simeq$ 1 for iron plasma composition)

\bigskip

Figure-4

\bigskip

Effect of finite-temperature ($T\simeq 5.9\times 10^6K$) on Fermi-Dirac plasma mass-density quantization for a chemical potential value, $\mu=100$. The oscillatory nature of the plasma quantities are smeared-out as the plasma temperature is increased.

\bigskip

Figure-5

\bigskip

Effect of the Coulomb instability in the absence of oscillatory part of plasma mass-density. The stability region (colored regions) up to very large values of relativity parameter, $x$, is bounded by critical field-parameter depending on the plasma composition. The field parameter values in these plots are slightly higher but consistent with the ones in Fig. 3. Also the left stability branches in these plots are due to approximation (ignoring the oscillatory part of the mass-density) which are not present in Fig. 3.

\bigskip

Figure-6

\bigskip

The solution to $y(\xi)$ for dense-core iron composition plasma with different values of magnetic field strengths. The solid and dashed curves indicate the stable and unstable configuration, respectively. The thickness of curves increases according to increase in the value of $\gamma$.

\bigskip

Figure-7

\bigskip

Effect of plasma composition and the strength of magnetic field on the compact-star mass-limit. The cases of $z=2,12,56$ shown respectively with curves of rectangular, filled-circles and halo-circles symbols.

\end{document}